# Visualizing nanoscale electronic band alignment at the $La_{2/3}Ca_{1/3}MnO_3$/Nb:$SrTiO_3$ interface

TeYu Chien(簡德宇)[1]*, Jian Liu[2], Jacques Chakhalian[2], Nathan P. Guisinger[3], & John W. Freeland[1]

[1]*Advanced Photon Source, Argonne National Laboratory, Argonne, IL 60439*

[2]*Department of Physics, University of Arkansas, Fayetteville, AR 72701*

[3]*Center for Nanoscale Materials, Argonne National Laboratory, Argonne, IL 60439*

Cross-sectional scanning tunnelling microscopy and spectroscopy (XSTM/S) were used to map out the band alignment across the complex oxide interface of $La_{2/3}Ca_{1/3}MnO_3$/Nb-doped $SrTiO_3$. By a controlled cross-sectional fracturing procedure, unit-cell high steps persist near the interface between the thin film and the substrate in the non-cleavable perovskite materials. The abrupt changes of the mechanical and electronic properties were visualized directly by XSTM/S. Using changes in the DOS as probe by STM, the electronic band alignment across the heterointerface was mapped out providing a new approach to directly measure the electronic properties at complex oxide interfaces.

* tchien@anl.gov

Recently, interfaces have become a playground for manipulation of strongly correlated electrons. The broken symmetry and modified local interactions have been shown to generate wholly new electronic phases not attainable in the bulk layers of a complex oxide heterostructure. For example, a superconducting phase was observed at the interface of insulating and metallic cuprate materials[1]; spin rearrangement and orbital reconstruction at the interface between a ferromagnetic manganite and superconducting cuprate[2,3]; metallic conductivity at the interface between polar and non-polar insulating perovskites[4,5]. However, to date our understanding of these interface driven phases is still limited. While tools exist to visualize the atomic structure, chemical state, and magnetic arrangement at the interface, the physical properties at the interfaces related to the charge carriers have thus far been only detected with non-local techniques. To achieve the next level of understanding and control of correlated electron behaviour, a nanoscale electronic probe with the requisite spatial resolution and sensitivity to changes in the local density of states (LDOS) near the highest occupied levels is clearly needed.

In investigating an interface, one of the key questions is how the electronic structure is altered crossing the boundary between the two materials. Following the extensive knowledge of semiconductor hetero-interfaces, the band-bending scenario is conventionally assumed to describe properties of oxide/oxide interfaces[6-9] and metal/oxide interfaces[10-12]. However, the applicability of this description can be questioned since in many instances these oxides are strongly correlated metals or Mott insulators which cannot be described within the framework of single-electron band theory. In fact, to date even though several theoretical efforts exist[13-17], it has not been experimentally established that the formalism used to describe semiconductors can be applied to strongly correlated interfaces. This is why a technique to directly visualize the local electronic properties at complex oxide interfaces is critical to understanding the physics at complex oxide interfaces.

More than a decade ago, cross-sectional scanning tunnelling microscopy (XSTM) was used to map the local electronic structure across semiconductor heterojunctions[18]. However, to the best of our knowledge, no analogous method was utilized to directly probe band bending across a buried oxide interface. Among others, the main challenge has been the lack of the natural cleavage plane in many complex oxide materials. Recently, Basletic *et al.* adopted conducting-tip atomic force microscopy (CT-AFM) to study the mechanically polished cross-section of epitaxial oxide interfaces, confirming the spatial variation of conductivity across the $LaAlO_3/SrTiO_3$ interface[19,20] inferred by Ohtomo *et al.*[4]. Due to the nature of CT-AFM, this approach is strictly limited to measuring the mobility of carriers at the Fermi level. To increase the spatial resolution and to provide a more detailed picture of the electronic structure, the approach of XSTM is well suited.

In this article, we have applied cross-sectional scanning microscopy and spectroscopy (XSTM/S) to investigate the local properties of a perovskite oxide heterojunction between ferromagnetic $La_{2/3}Ca_{1/3}MnO_3$ (LCMO) and n-type semiconducting Nb-doped STO (Nb:STO). The results reveal abrupt changes of topography and LDOS, and present a method to depict how the bands are altered across the interface. The conduction band across the LCMO and Nb:STO interface is shown to be aligned, which implies that the Fermi levels in both oxides are located roughly at the same energy level below the conduction band minimum (CBM). On the other hand, the valence bands are not aligned and any band bending near the interface was found to occur on a length scale less than 10 nm.

The 150nm-thick $La_{2/3}Ca_{1/3}MnO_3$ films were grown on 0.5mm-thick atomically flat (001) Nb:STO single crystal substrates using pulsed laser deposition. A stoichiometric $La_{2/3}Ca_{1/3}MnO_3$ ceramic target was ablated by a KrF excimer Laser (248 nm) with a 3 Hz repetition rate. The growth temperature was 730 °C and the oxygen partial pressure was 0.3 mbar. After deposition, samples were cooled down to 700 °C and annealed for half an hour at 0.5 bar oxygen pressure before cooling to room temperature to minimize the formation of oxygen vacancies.

After the thin film was grown on the Nb:STO substrate, the samples were cut half-way through the substrate with a precision dicing saw and clamped vertically on a sample holder. The sample was then fractured in ultra-high vacuum (UHV) at room temperature by applying a slow, steady force at the top portion of the sample above the cut. When the force reaches a critical strength, a crack initiates in the cut region and propagates across the sample.

All STM experiments were performed at room temperature with tungsten tips at a tunnelling set point of 20 pA at 3.0 V. The sample-tip polarity is defined as the sample relative to the tip, while the sample is grounded. The base pressure of the UHV chamber was below $10^{-10}$ torr. The data for Fig. 3 was measured with point-by-point dI/dV curves with the feedback "off" over a 300 nm × 300 nm grid with a 10 nm spacing between points starting at a set point of +3.0 V and 100 pA.

Figure 1 (a) shows the schematic of the atomic arrangement at the hetero-interface between LCMO and STO. Since both LCMO and STO have the perovskite $ABO_3$ structure, there is no cleavage plane, which poses a challenge to create an atomically flat surface in cross-section. Thus, the first key step in this approach is the generation of a suitably flat surface near the interface in cross-section when fracturing the system. Figure 1 (b)-(d) shows both optical microscopy and scanning electron microscopy (SEM) images of the fractured region of the LCMO/Nb:STO sample. The dimension of the fractured region is ~0.5 mm × 0.5 mm. The patterns of reflected light observed with an optical microscope (Fig. 1 (b)) indicate that the fractured surface is likely not globally flat, which is a signature of conchoidal fracturing common to fractured perovskite materials[21]. Figure 1 (c) shows the SEM image taken at the same magnification as the optical image. As seen in the image, a bright curved feature observed near the center of the fractured region, originating from the surface-tilt contrast effect due to the secondary electron yield[22], indicates that there is only one region with an apparent height change.

Our previous studies have shown that, even though this system does not have a cleavage plane, macroscopic size local regions with atomically flat surfaces can be created routinely by this fracturing procedure[23-25]. A topographic survey on one of the fractured samples allowed us to conclude that indeed most of the flat regions seen in SEM can be further assessed by STM[25]. To study the interface by STM, it is crucial to have a flat region at the vicinity of the interface, which locates at sample edge in the cross-sectional geometry. Figure 1 (d) shows a zoom-in view of the SEM image at the sample edge. As clearly seen, a well-defined region of brighter contrast and thickness of ~150 nm along the sample edge was observed and is indicated by an arrow. This region maintains its presence along the film side of the sample with uniform thickness and is due to the high-quality LCMO thin film.

Figure 2 (a) shows a 300 nm × 300 nm STM topographic image measured at the edge of the sample, near the red rectangular region shown in Fig. 1 (c); while Fig. 2 (b) depicts the dI/dV image recorded simultaneously at bias of +3.0 V. As shown in Fig. 2 (a) and (b), abrupt changes in surface topography and in electronic properties can be observed upon crossing the LCMO/Nb:STO interface. In the right half of the images (STO region), we found alternating SrO- and $TiO_2$-terminated surfaces as concluded from both the topography (Fig. 2 (a)) and the LDOS (Fig. 2 (b)). This result is consistent with our previous study of the fractured surfaces of bulk Nb:$SrTiO_3$[23]. On the other hand, on the left half of the images, which corresponds to the onset of the LCMO thin film, a dramatic difference in topography and in LDOS is clearly observed.

Topographically, unlike the STO region, no apparent order or atomic steps are seen in LCMO region. Instead, a rumpled surface following the trend of surface morphology in the STO region has appeared. The local roughness of the rumpled LCMO region is determined to be about one unit cell (0.4 nm) that can be compared to the local roughness of the SrO-terminated surface of the STO region of about half unit cell (0.2 nm)[23]. At the same time, electronically, a uniform contrast in the LDOS measured at +3.0 V (see Fig. 2 (b)) was found in the LCMO region along with a sharp transition at the interface. Since the edge of the sample can be determined unambiguously by detecting the sudden drop of the STM tip while scanning off the sample, the thickness of the LCMO region was determined to be ~150 nm in good agreement with the brighter contrast region seen in the SEM image (Fig. 2 (c)).

Presently, it is unknown what mechanism is precisely responsible for such a rumpled surface of the fractured LCMO, but this observation clearly indicates that fractured film and substrate may possess distinct surface morphologies. As a possible explanation consider the effect of strain. First, we note that since the film and substrate were fractured simultaneously, they experience the same strain (*i.e.* an iso-strain fracture). Because the Young's modulus of STO ($E$ = 238 GPa[26]) is about twice of that of LCMO ($E$ = 120 GPa[27]), locally STO is subjected to about twice the stress that is felt by LCMO. Furthermore, it is unlikely that the threshold stress for fracturing STO is also twice of that of LCMO. Thus, the thin LCMO layer (~150 nm thick) is likely fractured under the condition specific to the threshold of the substrate. Another important factor is that the LCMO layer is under ~1.1% tensile strain[28] which will in turn alter the stress-strain relationship. Based on these we may speculate that when creating the surface by fracturing, the exposed LCMO will likely try to relax and release this strain via the morphological instability resulting in the rougher surface.

In order to understand the local electronic structure, we investigated the LDOS by measuring the bias-dependent dI/dV curves. The positive (negative) bias part corresponds to the LDOS of conduction (valence) band with the Fermi level located at zero bias. Figure 2 (d) shows dI/dV measurements taken at points A, B and C in Fig. 2 (b), which represents the LCMO, $TiO_2$- and SrO-terminated Nb:STO regions, respectively. The valence band maximum (VBM) and conduction band minimum (CBM) can be extracted from the dI/dV curve by determining the initial upturning point for both negative and positive bias, as indicated in Fig. 2 (d). Previous work on fractured Nb:STO surfaces showed that applying a high sample-tip bias may lead to material transfer between the surface and the tip[24], which is why reliable dI/dV measurements were limited to the bias range from +3.0 V to -2.5 V. For the LCMO data, the position of VBM is consistent with the previous results from photoemission spectroscopy (PES)[29-32]. On the other hand, since the VBM of Nb:STO is located at about -3.2 V, the dI/dV measurements do not include the initial upturning point for Nb:STO at negative bias. The difference of dI/dV curves in the negative bias region provides an ideal marker for distinguishing between the two different oxides. Note that even though both systems are metallic, low carrier density implies that states close to the Fermi level have low spectral weight and are hard to probe with the low tunnelling currents used here for the LDOS measurements.

To electronically identify the interface, the negative bias measurements are essential but present an experimental challenge. Because of the low valence band LDOS in Nb:STO under negative bias conditions, the imaging mode is not stable. In order to circumvent this issue, we measured point-by-point dI/dV curves with the feedback “off”. Based on this three-dimensional data set, we were able to generate LDOS images at any bias condition. As noted above, dI/dV at negative bias can be used as an ideal marker, here we choose to generate a LDOS(dI/dV) mapping at the bias (-2.5 V) where the maximum contrast in dI/dV curves between two materials were observed and, as shown in Fig. 3, the LCMO is clearly seen as a region of brighter contrast. This method provides an ideal way to electronically determine the interface between Nb:STO and LCMO via a valence band contrast. More generally, using this approach, one can image interface between dissimilar materials with atomic resolution utilizing valence or conduction band contrast.

In addition to providing a contrast mechanism to image the film-substrate interface, the data for the LDOS map can be used to extract a direct picture of how the bands evolve across the interface. By averaging the dI/dV curves along the lines parallel to the interface, we plot the changes in the averaged dI/dV curves on crossing the LCMO/Nb:STO interface (see Fig. 4 (a)). As clearly seen, the local dI/dV curves with positive bias in both materials are almost identical even for the curves measured closest to the interface. In contrast, a significant change in the valence band is found between LCMO and Nb:STO. Figure 4 (b) shows the color plot of the averaged dI/dV curves of the same data upon crossing the interface. The CBM and VBM in Fig. 4 (b) allow us to track the band structure across the LCMO/Nb:STO interface. In a band bending scenario, one would expect to see the CBM and VBM change when approaching the interface. Upon direct examination, it is revealed that the CBM on crossing the interface stays unchanged, implying that any band bending across the interface is confined to the spatial resolution of $<10$ nm. On the other hand, a significant change in the VBM is found between LCMO and Nb:STO. By examining the VBM in LCMO, within the error we find no noticeable change on approaching the interface, thus confirming that the bending is confined to a length scale smaller than 10 nm. This finding agrees well with the commonly cited length scale for LCMO/STO interface[6]. However, our direct band mapping results presented here is clearly different from the conventional band model used for explaining the I-V properties of the LCMO/Nb:STO junction[33], in which the conduction bands are not aligned. At the same time we cannot exclude that the exposed interface subjected to strain-relaxation might raise extra surface states or surface defect states, which could in turn pin the Fermi levels across the interface leading to the observed differences with electrical transportation measurements. First-principle calculations for the exposed cross-sectional interface with strain relaxation system would be necessary to clarify the discrepancy. It is worth to note that the poor spatial resolution mainly comes from the deterioration of the tip from debris on the fractured surface. Future efforts are focused on improving the surface quality without changing the properties.

Another interesting observation is related to the possible charge transfer across the interface. We recap, that at room temperature, the conducting state of Nb:STO is induced by Nb doping, while pure STO is a band insulator[34]. As seen in Fig. 2 (d), the Fermi level of the Nb:STO is located closer to the CBM than to the VBM, which indicates that the Nb is donating electrons to the conduction band (*n*-type). On the other hand, LCMO at this composition is a *p*-type material. The charge transfer between these two materials across the *p*-*n* junction as well as band bending can be expected. To resolve the details of the charge transfer across the interface, higher spatial resolution than presented here is required, and is currently under investigation.

In conclusion, a novel application of XSTM was successfully utilized to reveal the topographic change across the complex oxide interface. In addition, LDOS probed by XSTS revealed that the CBM in both materials are well aligned, and an abrupt change is observed only in the VBM. These findings highlight the vital importance of locally probed nanoscale electronic properties in the vicinity of an oxide-oxide interface. This XSTM based tool provides a new doorway to explore physics of correlated carriers in complex oxide heterojunctions. In the future, one can envision simultaneously monitoring the band dynamics as function of bias applied across the oxide interfaces, direct detection of the local magnetic properties with spin polarized STM tips, or visualizing regions of interfacial superconductivity or high-mobility conduction at the interface.

Authors acknowledge the valuable discussion with Dr. T.S. Santos. Work at Argonne, including the Center for Nanoscale Materials, is supported by the U.S. Department of Energy, Office of Science, Office of Basic Energy Sciences, under Contract No. DE-AC02-06CH11357. J.C. and J.L. were supported by NSF Grant No. DMR-0747808.

Figure Captions:

Figure 1: (Color online) (a) The schematic atomic arrangement near the hetero-interface of LCMO and STO. (b) Optical microscopy image of fractured LCMO/Nb:STO surface. (c) SEM image with same scale as in (b). (d) A magnified SEM image, which reveals contrast between thin film (LCMO) and substrate (Nb:STO).

Figure 2: (Color online) (a) Topography measured by STM at sample edge. (b) dI/dV recorded simultaneously with the topography with set point as 3.0 V; 20 pA. (c) SEM image with the same scale at sample edge. (d) dI/dV curves measured at points A, B and C, indicated in (b), on LCMO, $TiO_2$- and SrO-terminated Nb:STO, respectively.

Figure 3: (Color online) LDOS mapping at -2.5 V distinguishes LCMO and Nb:STO electronically.

Figure 4: (Color online) (a) averaged dI/dV curves measured along from LCMO crossing the interface into Nb:STO, as indicated with an arrow in Fig. 3; (b) Experimental band diagram across LCMO/Nb:STO interface. The Fermi energy is indicated as the dashed line at zero bias. The CBM and VBM were obtained from each individual dI/dV curve. The average values and errors are marked as open red circles and error bars, respectively, connected with lines, which shows the band mapping across the LCMO/Nb:STO interface.

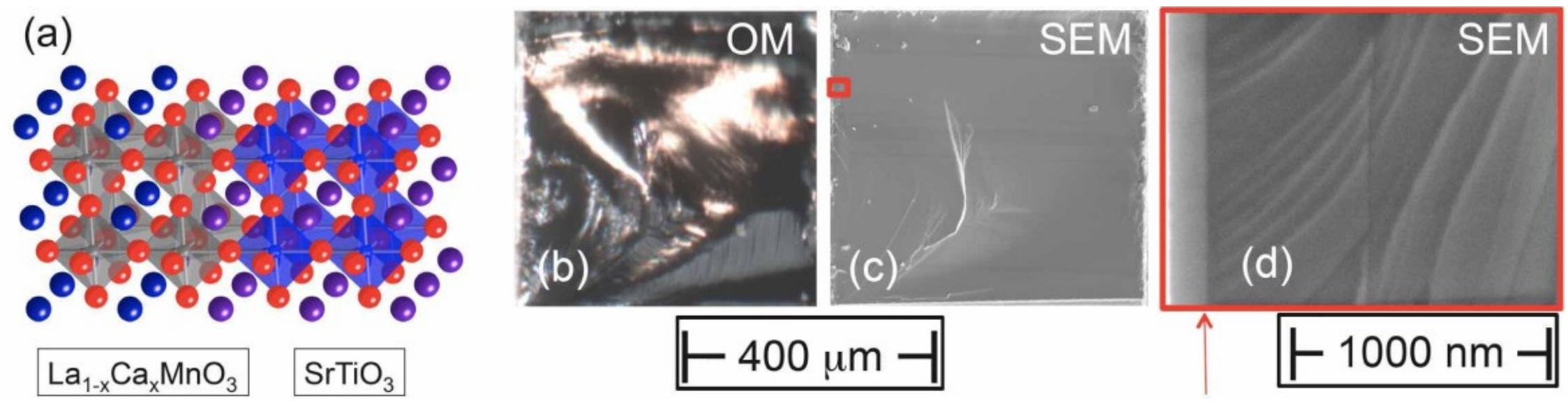

Figure 1

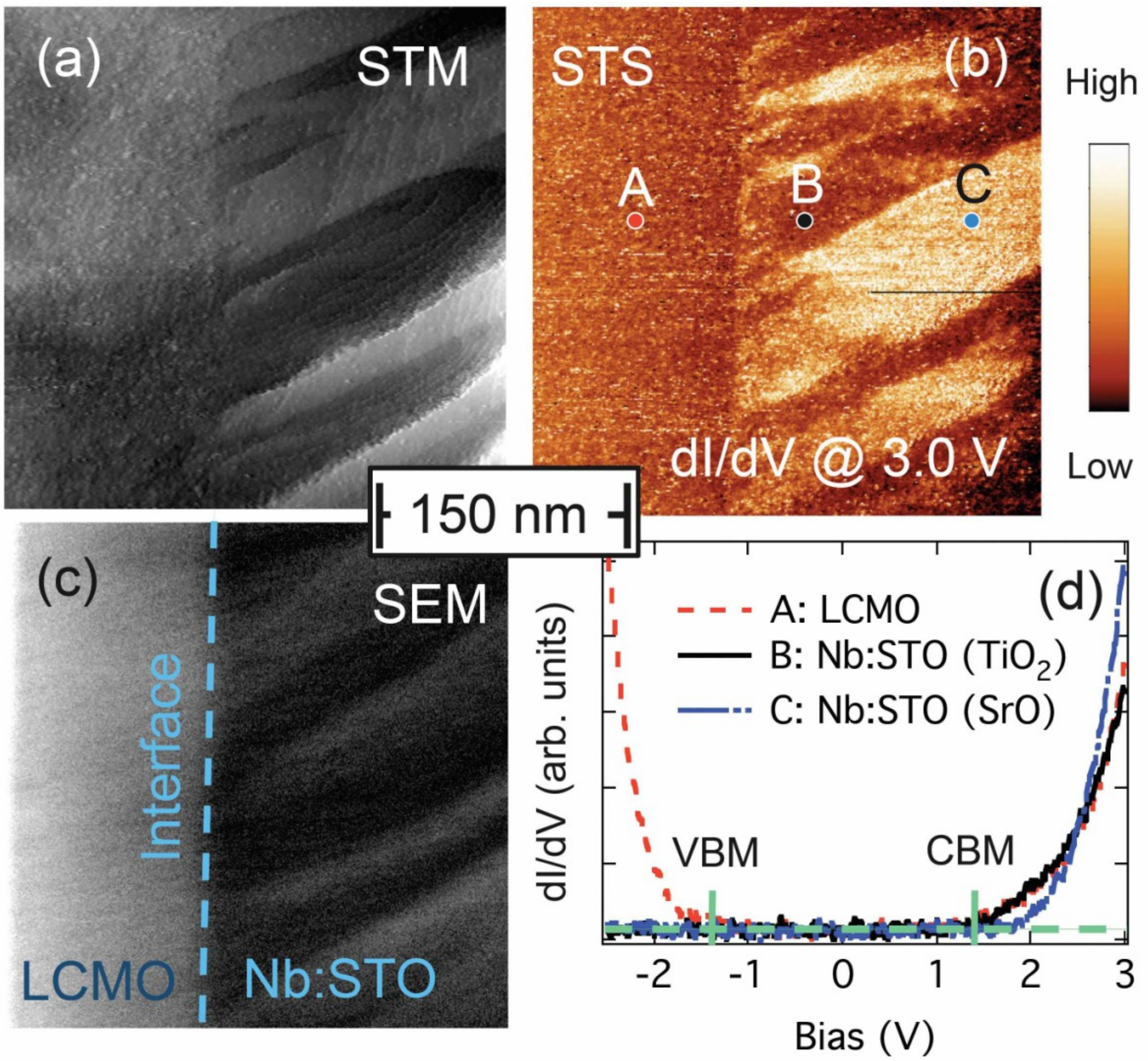

Figure 2

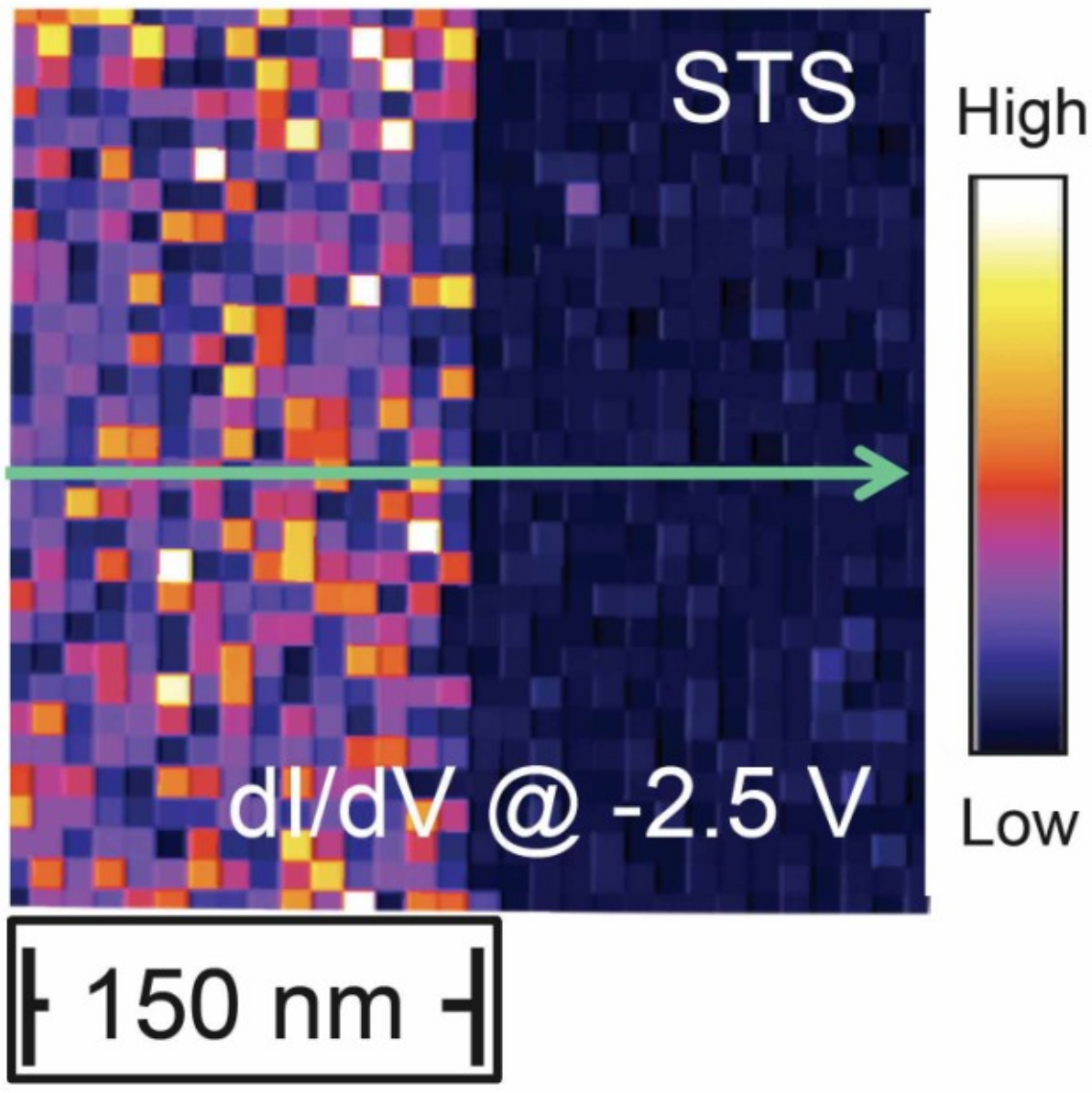

Figure 3

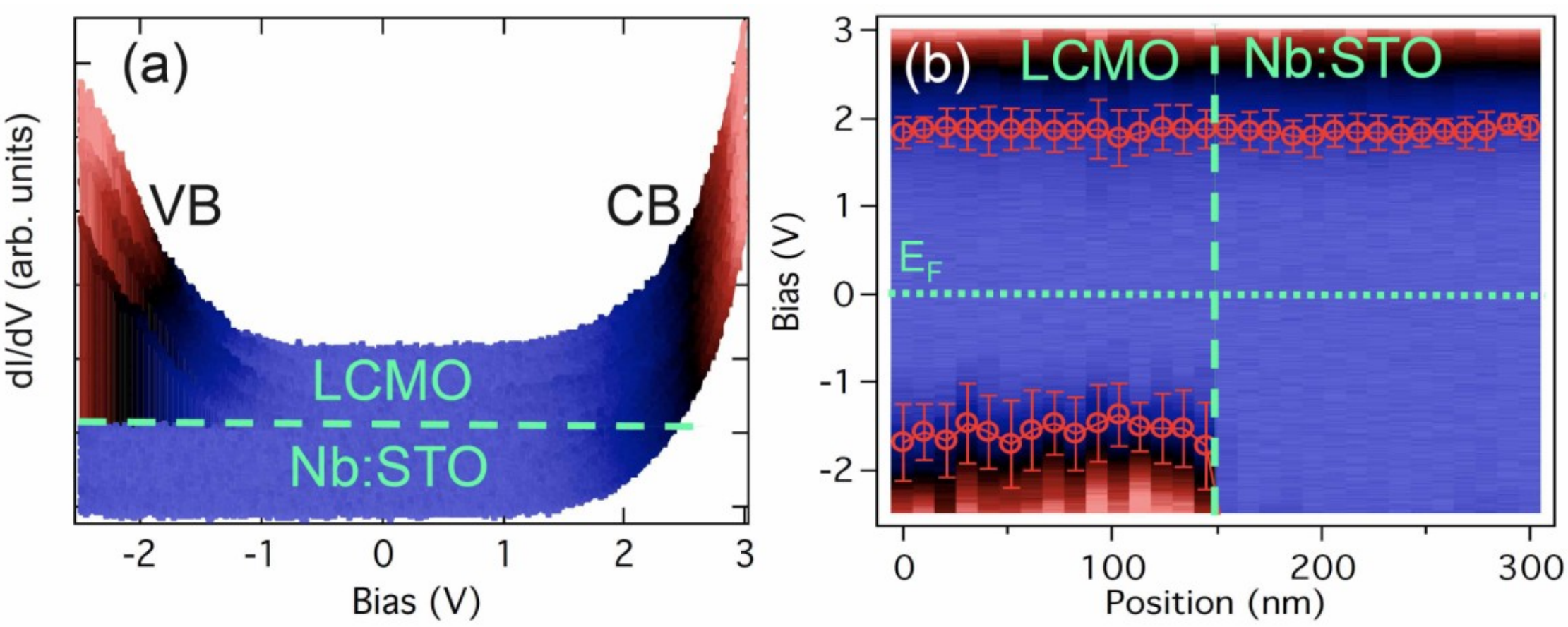

Figure 4